\documentclass[aps,pra,reprint,superscriptaddress,floatfix,showpacs,showkeys]{revtex4-2}
\usepackage{xcolor}
\usepackage{cancel}
\usepackage{float}
\usepackage{lipsum}
\usepackage{amsmath}
\usepackage{amssymb}
\usepackage{empheq} 
\usepackage{braket}

\usepackage{bm}
\usepackage{graphicx}
\usepackage{hyperref}
\usepackage{siunitx}
\usepackage{multibib}
\hypersetup{
    colorlinks,
    linkcolor={blue!50!blue},
    citecolor={blue!50!blue},
    urlcolor={blue!
    80!black}
}
\newcites{SuppRefs}{Supplementary References}

\usepackage{filecontents}

\begin{document}

\title[High temperature mid-IR polarizer via natural in-plane hyperbolic Van der Waals crystals]{High temperature mid-IR polarizer via natural in-plane hyperbolic Van der Waals crystals}
\author{Nihar Ranjan Sahoo}
\thanks{These two authors contributed equally}
\author{Saurabh Dixit}
\thanks{These two authors contributed equally}
\author{Anuj Kumar Singh}
\affiliation{Laboratory of Optics of Quantum Materials, Physics Department, Indian Institute of Technology Bombay, Mumbai - 400076, India}
\author{Sang Hoon Nam}
\author{Nicholas X. Fang}
\email{nicfang@mit.edu}
\affiliation{Department of Mechanical Engineering, Massachusetts Institute of Technology, Cambridge, MA-02139, USA}
\author{Anshuman Kumar}
\email{anshuman.kumar@iitb.ac.in}
\affiliation{Laboratory of Optics of Quantum Materials, Physics Department, Indian Institute of Technology Bombay, Mumbai - 400076, India}
\date{\today}

\begin{abstract}
Integration of conventional mid to long-wavelength infrared polarizers with chip-scale platforms is restricted by their bulky size and complex fabrication. Van der Waals materials based polarizer can address these challenges due to its non-lithographic fabrication, ease of integration with chip-scale platforms, and room temperature operation. In the present work, mid-IR optical response of the sub-wavelength thin films of $\alpha$-MoO$_3$ is investigated for application towards high temperature mid-IR transmission and reflection type thin film polarizer. To our knowledge, this is the first report of above room temperature mid-IR optical response of $\alpha$-MoO$_3$ to determine the thermal stability of the proposed device. We find that our $\alpha$-MoO$_3$ based polarizer retains high extinction ratio with peak value exceeding 10 dB, up to a temperature of 140$^{\circ}$C. We explain our experimental findings by natural in-plane hyperbolic anisotropy of $\alpha$-MoO$_3$ in the mid-IR, high temperature X-ray diffraction and Raman spectroscopic measurements. This work opens up new avenues for naturally in-plane hyperbolic van der Waals thin-films to realize sub-wavelength IR optical components without lithographic constraints.
\end{abstract}

\keywords{Hyperbolic in-plane anisotropy, natural hyperbolic materials, van der Waals crystal, 2D materials, mid-IR polarizer}

\maketitle

\section{Introduction}
Recently, a new class of van der Waals (vdW) layered materials have been shown to possess natural hyperbolic anisotropy in mid-infrared (IR) spectral region\cite{Ma2018,TaboadaGutirrez2020,Caldwell2019,Biswas2021}. Hyperbolic anisotropy is an extreme type of optical anisotropy in which the real part of dielectric permittivity holds the opposite sign in different crystallographic directions. As a result, within a hyperbolic spectral region, the vdW material behaves like a metal in one crystal direction and a dielectric in the other crystal direction. Unlike artificial hyperbolic metamaterials\cite{Poddubny2013}-- where in-plane hyperbolic anisotropy is invoked by lithographic patterning, natural hyperbolicity of these vdW materials is attributed to structural anisotropy of crystal unit cell. In particular, h-BN\cite{Caldwell2019}, $\alpha$-MoO$_3$\cite{Ma2018,Zheng2019} and $\alpha$-V$_2$O$_5$\cite{TaboadaGutirrez2020} are natural hyperbolic materials (NHMs) which exhibit Restrahalen Bands (RBs)-- the spectral region between longitudinal optical (LO) and transverse optical (TO) phonons - in the mid-IR spectral region and show hyperbolicity due to interaction of optical phonons with photons (light-matter interaction). Phonons have a relatively long lifetime compared to plasmons resulting in lower optical losses than their analogous plasmonic-based metamaterials\cite{Xia2014,Chang2016,Wang2020}$\-$ in which photons are coupled with plasmons. Many NHMs \cite{Ma2018,TaboadaGutirrez2020,Caldwell2019} exhibit hyperbolic anisotropy in mid-IR spectral region (3$\mu$m -  30$\mu$m) which has diverse applications like polarized IR imaging\cite{Tong2020}, molecular sensing\cite{Desouky2018,Yoo2020}, free space communication\cite{Pirotta2021} and quantum interference\cite{nalabothula2020engineering}. Unlike h-BN, which possesses uniaxial hyperbolic anisotropy (i.e. $\varepsilon_{xx}$ = $\varepsilon_{yy}$ $\neq$ $\varepsilon_{zz}$), $\alpha$-MoO$_3$ exhibits in-plane hyperbolic anisotropy (i.e. $\varepsilon_{xx}$ $\neq$ $\varepsilon_{yy}$ $\neq$ $\varepsilon_{zz}$) which is particularly beneficial for planar mid-IR optical devices\cite{AbediniDereshgi2020,Ou2020}. With this motivation, there has been recent interest in developing flat optics based on vdW layered materials which can be integrated with chip-scale platforms using vdW integration, operational at room temperature, and does not involve complex lithographic fabrication techniques\cite{Caldwell2016,Kim2019,Guo2017,Law2013}.\par
\begin{figure*}
    \includegraphics[width=\textwidth]{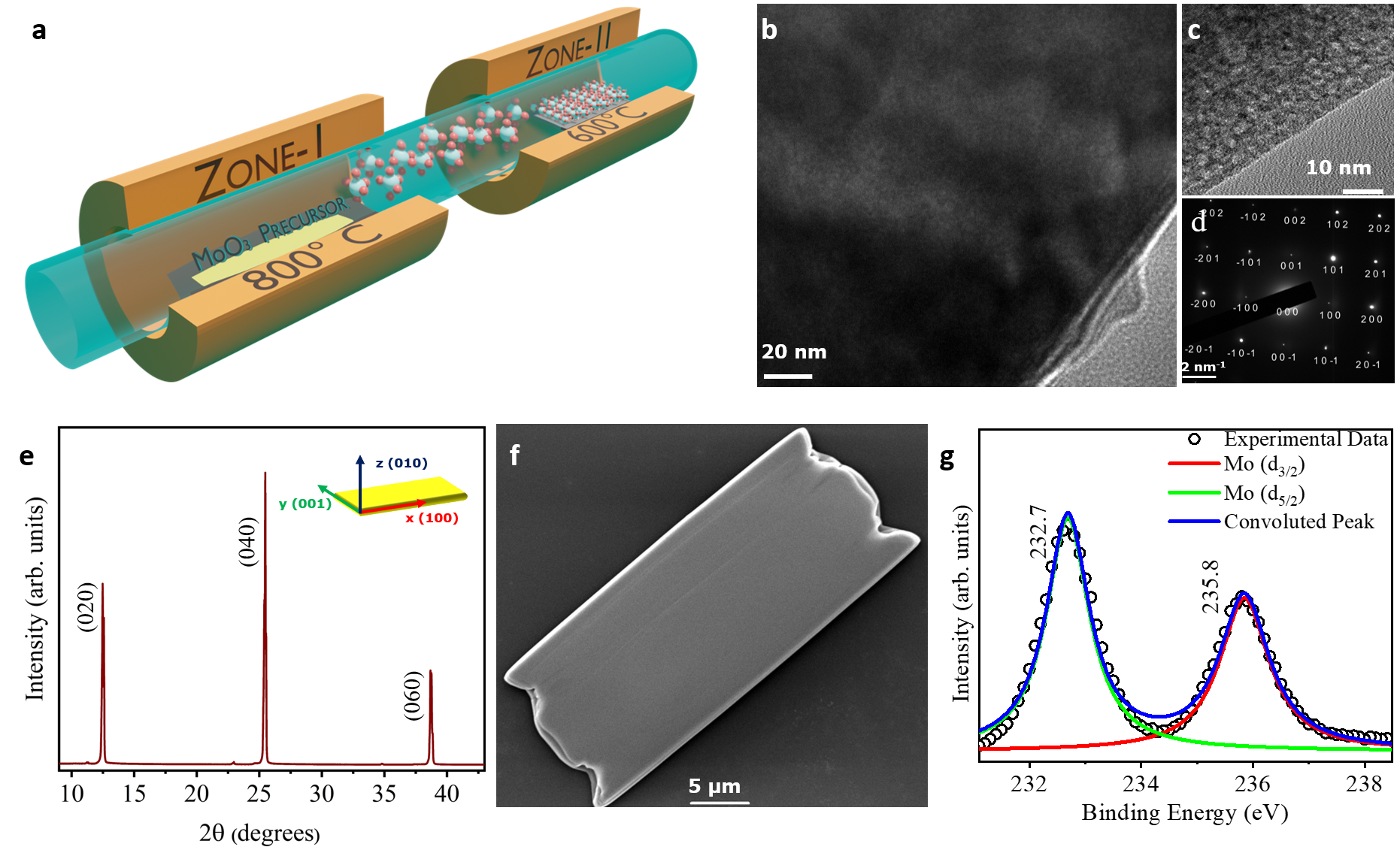}
    \caption{\textbf{Structural and chemical properties of fabricated $\alpha$-MoO$\textsubscript{3}$ thin films.} (a) Schematic illustration of thermal physical deposition technique for fabrication of $\alpha$-MoO$\textsubscript{3}$ thin films. (b)-(c) High-resolution transmission electron microscopic images and (d) selected area electron diffraction pattern of the fabricated $\alpha$-MoO$\textsubscript{3}$ thin film. (e) X-ray Diffraction of $\alpha$-MoO$\textsubscript{3}$ thin films (inset shows the crystallographic direction of $\alpha$-MoO$\textsubscript{3}$ planes with the coordinate axes). (f) A representative scanning electron microscopic image of as-deposited $\alpha$-MoO$\textsubscript{3}$ thin films on a silicon substrate. (g) X-ray Photoelectron Spectroscopy of $\alpha$-MoO$\textsubscript{3}$ thin film where the scattered plots represent the experimental data and solid lines are the fitted data for the binding energies of Mo d\textsubscript{5/2} and Mo d\textsubscript{3/2}. }
\end{figure*}

We present an application of $\alpha$-MoO$_3$ in the mid-IR spectral region, i.e. from 545$^{-1}$ - 1000 cm$^{-1}$ (around 10$\mu$m - 18$\mu$m), as a thin film polarizer that reflects the light with one state of polarization while transmitting the light with its orthogonal state of polarization. Here, single-crystal $\alpha$-MoO$_3$ thin films are synthesized using physical vapor deposition (schematically shown in Fig.~1(a)) and are transferred on top of potassium bromide (KBr) window, purchased from Edmund Optics, using mechanical exfoliation technique. In-plane anisotropy of the synthesized $\alpha$-MoO$_3$ thin film is confirmed using polarization-resolved Raman spectroscopy. We optimize the mid-IR optical responses of $\alpha$-MoO$_3$, mainly transmittance and reflectance, as a function of the thickness. Optimum thickness of $\alpha$-MoO$_3$ based IR polarizer is found to lie in the range of 2.5$\mu$m - 3.5$\mu$m for which the extinction ratio (ER) is obtained more than 7.5 dB and 10 dB in a broad mid-IR spectral region, respectively for reflection and transmission geometry with remarkable operational bandwidth and degree of polarization (DOP) in long-wavelength IR regime.  Moreover, to the best of our knowledge, this is the first work where the optical response of $\alpha$-MoO$_3$ is studied above room temperature, to determine the operating temperature tolerance for the $\alpha$-MoO$_3$ thin-film based polarizer device. We observe that the ER and DOP of $\alpha$-MoO$_3$ thin films retain their values up to a temperature of 140 $^{\circ}$C. We explain these experimental results through extensive full wave simulations of Maxwell's equations and via complementary measurements using temperature dependent X-ray diffraction and Raman spectroscopy.\par

\begin{figure*}
    \includegraphics[width=\textwidth]{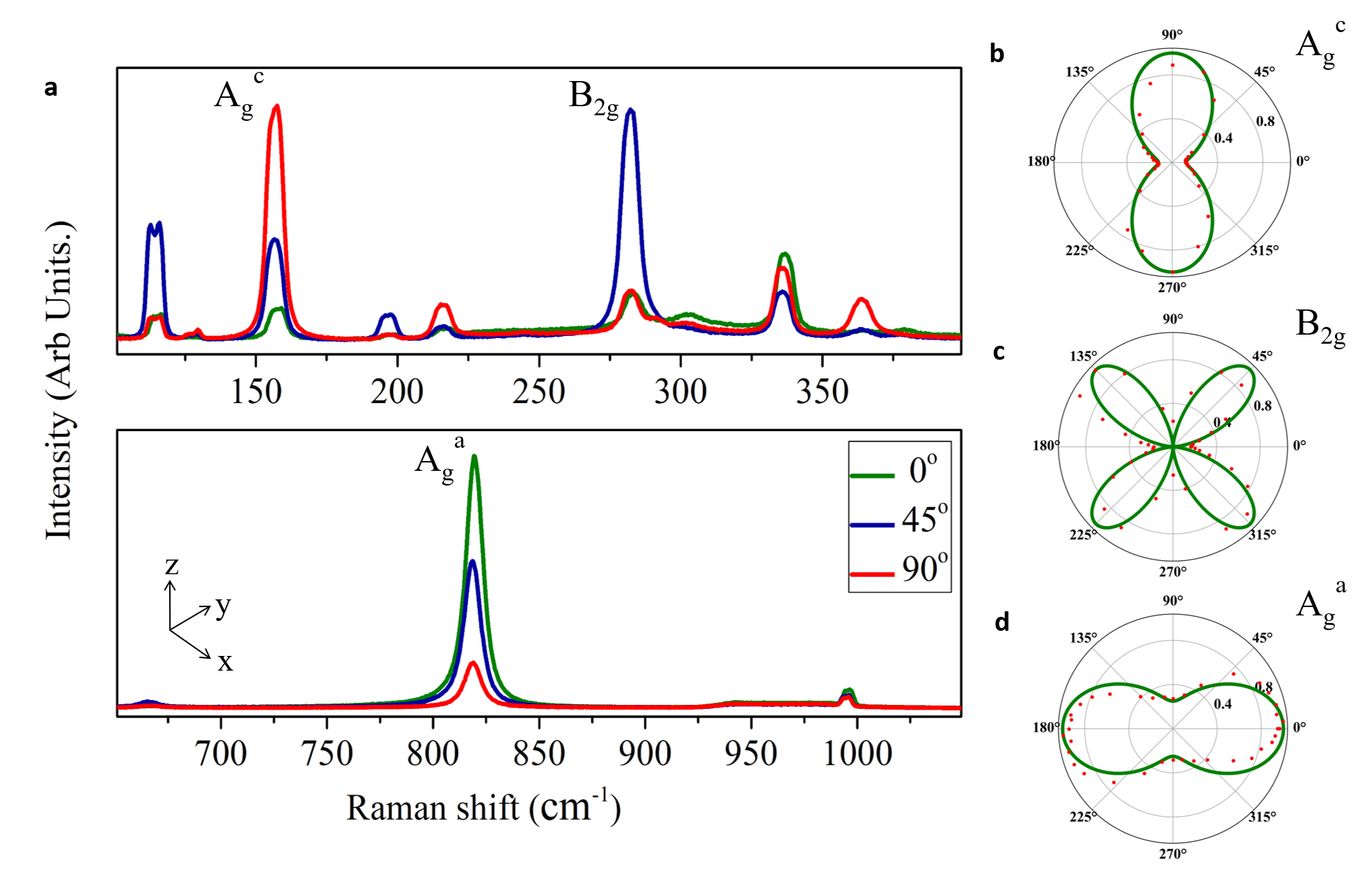}
    \caption{\textbf{Polarization resolved Raman scattering spectroscopy.} (a) Polarization resolved raman spectra of $\alpha$-MoO$\textsubscript{3}$ thin film at $\beta$ $\sim$ 0 $^{\circ}$ (green), 45$^{\circ}$ (blue) and 90$^{\circ}$ (red) exhibiting characteristic phonon energies of $\alpha$-MoO$\textsubscript{3}$. Polar plot of angle resolved normalized Raman intensities for (b) A$\textsubscript{g}\textsuperscript{c}$, (c) B$\textsubscript{2g}$ and (d) A$\textsubscript{g}\textsuperscript{a}$ modes of $\alpha$-MoO$\textsubscript{3}$.}
\end{figure*}

\section{Results and Discussion}
We first confirm the structure of our fabricated crystals. Representative high-resolution transmission electron microscopy (HRTEM) and selected area electron diffraction (SAED) analysis of synthesized thin films are shown in Fig.~1(b)-(d). The estimated values of in-plane lattice constants are around 3.69 $\text{\AA}$ and 3.97 $\text{\AA}$. Further structural information of synthesized thin films is investigated via X-Ray diffraction pattern (shown in Fig.~1(e)) which shows sharp diffraction peaks at the 2$\theta$ values of 12.48$^{\circ}$, 25.48$^{\circ}$ and 38.76$^{\circ}$ corresponding respectively to (020), (040), and (060) planes of $\alpha$-MoO$\textsubscript{3}$ \cite{wang2017growth}. Out of plane lattice constant is found around 14.14 $\text{\AA}$ using Bragg's Law\cite{cullity1956elements}. The values of lattice constants evaluated from XRD, HRTEM, and SAED match closely with the lattice parameters of the orthorhombic phase of MoO$\textsubscript{3}$ (i.e. $\alpha$-MoO$\textsubscript{3}$)\cite{Ma2018,wang2017growth}. SEM image of the synthesized $\alpha$-MoO$\textsubscript{3}$ thin films on the silicon substrate is shown in Fig.~1(f), which reveals the rectangular shape of flake. Chemical composition of the thin films is examined using X-ray photoelectron spectroscopy. XPS spectrum exhibits binding energies of Mo 3d doublets at 232.7 eV (for 3d\textsubscript{5/2}) and 235.8 eV (for 3d\textsubscript{3/2}) which is associated with the VI oxidation state of molybdenum (Fig.~1(g)). These peaks indicate negligible oxygen vacancies in the synthesized $\alpha$-MoO$\textsubscript{3}$ thin films\cite{Liao2019}. These investigations substantiate the fact that the chemical and structural properties of our fabricated rectangular-shaped $\alpha$-MoO$\textsubscript{3}$ thin films using thermal physical deposition technique are as expected. 

Next, we explore the anisotropy of $\alpha$-MoO$\textsubscript{3}$ thin films using polarization-resolved Raman spectroscopy. The $\alpha$-MoO$\textsubscript{3}$ unit cell follows the D$^{16}_{2h}$ space group \cite{Wen2020} which consists of 4 Mo and 12 O (total 16 atoms). Hence, there are 48 phonon modes at the center of the Brillouin zone having the following decomposition\cite{Seguin1995}:
$$
\begin{aligned}
\Gamma^{\text {vib }} = & 8 A_{g} \oplus 4 B_{1 g} \oplus 4 B_{2 g} \oplus 8 B_{3 g} \oplus 4 A_{u} \\
& \oplus 7 B_{1 u} \oplus 7 B_{2 u} \oplus 3 B_{3 u}
\end{aligned}
$$

Out of these modes, A$_{g}$, B$_{1 g}$,  B$_{2 g}$, and B$_{3 g}$ are Raman active modes. Since $\alpha$-MoO$\textsubscript{3}$ has an orthorhombic crystal structure, it has different Raman active modes associated with phonons along different crystallographic directions. Here, we focus on the B$_{2 g}$ mode (at 282 cm$\textsuperscript{-1}$) which is the wagging mode of O-Mo-O atoms, A$_{g}$ mode for the translational vibration of MoO$_6$ chain along the $y-$ crystal direction (called A$_{g}^c$ mode) at 157 cm$\textsuperscript{-1}$, and A$_{g}$ mode from asymmetric stretching of O-Mo-O atoms along the $x-$ crystal direction is at 818  cm$\textsuperscript{-1}$ (called  A$_{g}^a$ mode)\cite{R3}. If e$_i$ and e$_s$ are the polarization unit vectors along the incident and scattered light, then Raman intensity is given by $I \propto\left|e_{\mathrm{i}} \cdot R \cdot e_{\mathrm{s}}\right|^{2}$. The angular dependence of the Raman intensity for A$_g$ and B$_{2g}$ modes can be obtained from:
$$
\begin{array}{l}
R\left(A_{g}\right)=\left(\begin{array}{ccc}
A & 0 & 0 \\
0 & B & 0 \\
0 & 0 & C
\end{array}\right)
\end{array}
$$
$$
\begin{array}{l}
R\left(B_{2g}\right)=\left(\begin{array}{ccc}
0 & 0 & E \\
0 & 0 & 0 \\
E & 0 & 0
\end{array}\right)
\end{array}
$$
Here, $A$, $B$, $C$ and $E$ represent the strength of Raman tensor's elements. Intensity of these Raman modes can be theoretically captured via the Raman tensor\cite{Wen2020} as:
\begin{equation}
\begin{array}{l}
I\left(A_{g}\right) \propto\left(A \cos ^{2} \beta + B \sin ^{2} \beta\right)^{2}\\
\\
I\left(B_{2 g}\right) \propto E^{2} \sin ^{2} 2 \beta
\end{array}
\label{eq:raman_I}
\end{equation}

Here, $\beta$ represents the angle between the [100] crystal direction and the polarization state of laser. In Fig.~2(a), our experimental Raman spectra show A$_{g}^c$, B$_{2 g}$ and A$_{g}^a$ modes at 156 cm$\textsuperscript{-1}$, 282 cm$\textsuperscript{-1}$, and 818 cm$\textsuperscript{-1}$ respectively for $\beta$ $\sim$ 0 $^{\circ}$, 45 $^{\circ}$ and 90 $^{\circ}$. These characteristic Raman modes of $\alpha$-MoO$\textsubscript{3}$ show a variation in the Raman intensities as a function of $\beta$, which is attributed to the dependence of Raman intensity tensor on the polarization state of incident laser light and the crystallographic directions. Fig.~2(b)-(d) present polar plots for theoretical and experimental Raman intensities of A$_{g}^c$, B$_{2 g}$ and A$_{g}^a$ respectively. Theoretical variation of Raman intensities of these modes is calculated using Eq. 1. B$_{2 g}$ mode has the angular periodic intensity variation of 90 $^{\circ}$ (four lobes), and the orientation of two-lobed A$_{g}$ modes are orthogonal to each other. For A$_{g}^c$ and A$_{g}^a$ mode, the ratio of $A$ and $B$ are taken as 0.35 and 2 in the Eq.\ref{eq:raman_I}, respectively, to fit our experimental polarization-resolved Raman data. Here, the condition $A>B$ results in the formation of a two-lobed main axis parallel to the $x$-axis (i.e. [100] direction). On the other hand, $A < B$ leads to a two-lobed main axis perpendicular to the $x$-axis. These results validate that synthesized $\alpha$-MoO$_{3}$ flakes exhibit the expected crystallographic orientation with strong in-plane anisotropy, which is crucial for the functioning of our proposed IR polarizer.\par

The optical response of a thin film depends on its thickness and hence it is a vital parameter for the polarizer. Therefore, we consider different thicknesses of $\alpha$-MoO$\textsubscript{3}$ thin film from the flakes transferred using mechanical exfoliation technique and investigate their mid-IR optical responses. Thicknesses of $\alpha$-MoO$\textsubscript{3}$ thin films are measured using a profilometer and are shown in Fig.~S2 of the supplementary information for three representative samples D1, D2, and D3, respectively. The mechanical exfoliation technique results in the transfer of vdW flakes with non-uniform thicknesses. For instance for samples D1, D2 and D3 thicknesses are found to be in the range 2.27-3.08 $\mu$m, 2.64-3.53 $\mu$m and 3.95-4.59 $\mu$m respectively.\par 

Next, we explore the mid-IR optical responses of $\alpha$-MoO$\textsubscript{3}$ thin films. $\alpha$-MoO$_3$ exhibits three RBs in the spectral range of 544.6 cm$\textsuperscript{-1}$ to 850.1 cm$\textsuperscript{-1}$ (RB-1), 821.4 cm$\textsuperscript{-1}$ to 963 cm$\textsuperscript{-1}$ (RB-2), and 956.7 cm$\textsuperscript{-1}$ to 1006.9 cm$\textsuperscript{-1}$ (RB-3), where the real part of dielectric permittivity is negative along [001], [100] and [010] crystal direction of $\alpha$-MoO$_3$ respectively\cite{alvarez2020infrared}. Due to the in-plane hyperbolic anisotropy of $\alpha$-MoO$\textsubscript{3}$ along [100] and [001] crystallographic directions in the RBs, $\alpha$-MoO$_3$ reflects light which is linearly polarized along one direction. In contrast, the light with an orthogonal polarization state gets transmitted as shown schematically in Fig.~3(a). Hence, $\alpha$-MoO$_3$ exhibits anisotropic reflectance and transmittance for $s-$ and $p-$ polarized light in the RB-1 and RB-2, respectively, making it a suitable candidate for reflecting as well as transmitting type mid-IR polarizer.\par

\begin{figure*}
    \includegraphics[width=\textwidth]{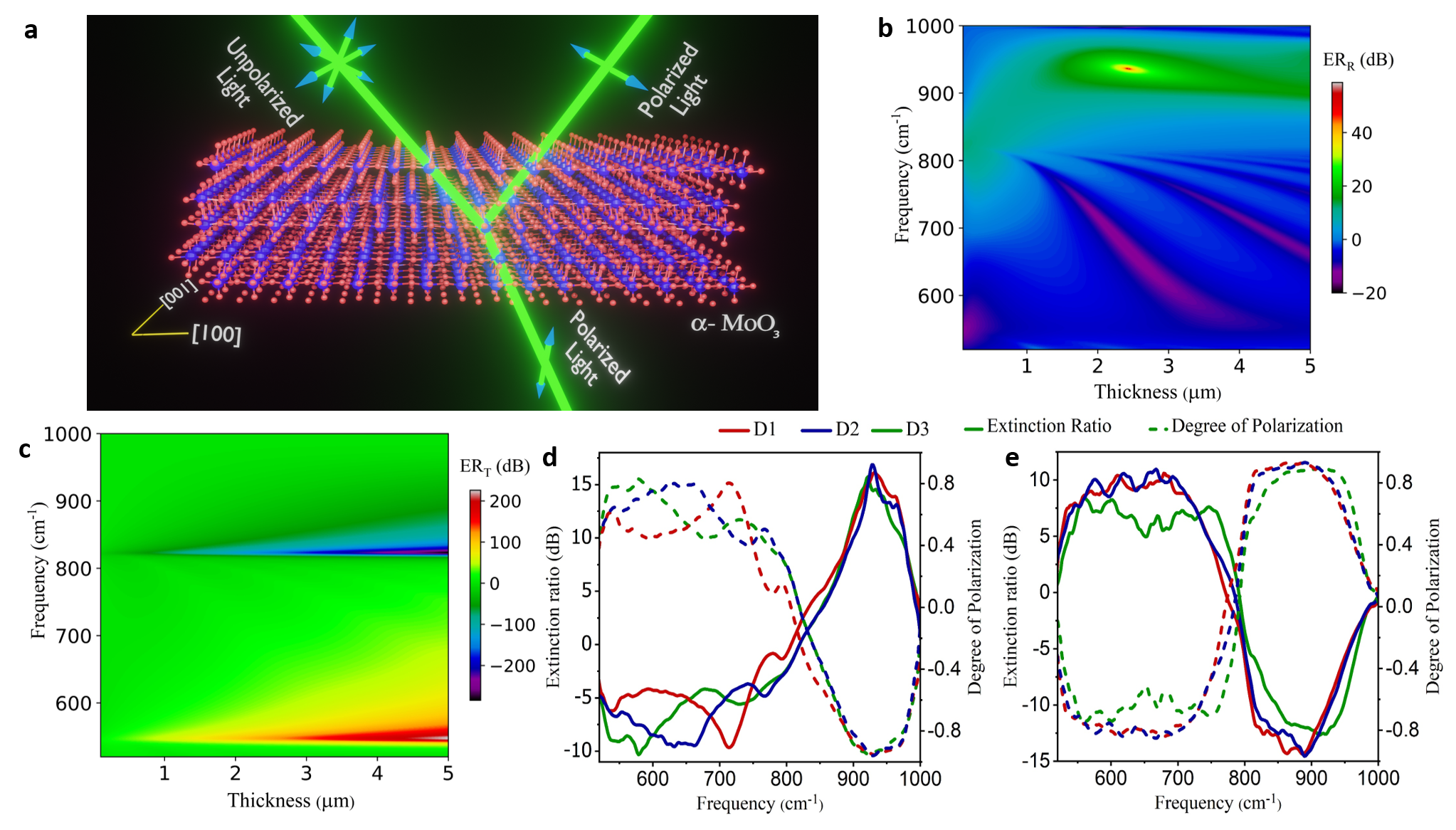}
    \caption{\textbf{Figures of merit of mid-IR polarizer in the transmission and reflection geometry.} (a) Schematic illustration of $\alpha$-MoO$\textsubscript{3}$ thin film as a reflecting and transmitting type polarizer device. (b) - (c) shows theoretical ER as a function of thickness and frequency for reflected and transmitted light respectively. (d) - (e) represent the ER and DOP of IR polarizer based on $\alpha$-MoO$\textsubscript{3}$ thin films in reflection and transmission geometry, respectively, for Samples D1 - D3. Solid lines and dashed lines correspond to ER and DOP respectively.}
\end{figure*}

\begin{table*}
\centering
\setlength{\tabcolsep}{10pt}
\renewcommand{\arraystretch}{1.5}
\begin{tabular}{|m{1.0cm} | m{2.0 cm}| m{2.0cm}| m{2.0cm}| m{2.0cm}|}
\hline
Samples & \multicolumn{2}{c|}{Reflection Bandwidth (cm$^{-1}$)}  & \multicolumn{2}{c|}{Transmission Bandwidth (cm$^{-1}$)}  \\
\cline{2-5}
 & RB-1  & RB-2 & RB-1 & RB-2 \\
\hline
D1 & 694 - 731 & 892- 977 & 545 - 726 & 810 - 924\\

D2 & 571 - 678 & 900 - 977 & 545 - 726 & 815 - 928\\

D3 & 545 - 613 & 892 - 972 & 586 - 608 & 847 - 948\\

\hline
\end{tabular}
\caption{\label{tab:example} Operational bandwidth for Samples D1 - D3 in both RBs of $\alpha$-MoO$_3$ in reflection and transmission geometry.}
\end{table*}

\begin{figure*}
    \includegraphics[width=\textwidth]{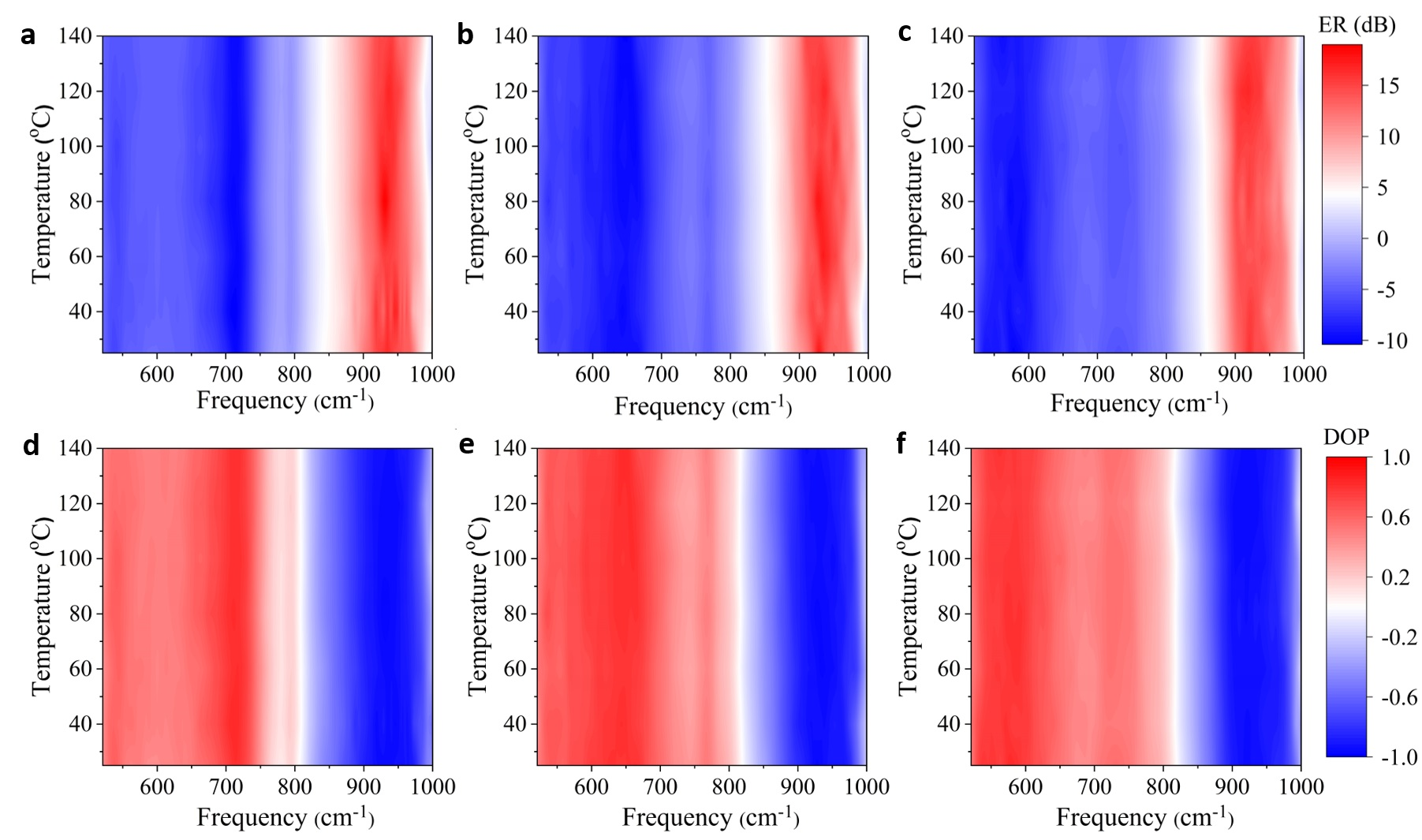}
    \caption{\textbf{Temperature-dependent figures of merit of $\alpha$-MoO$\textsubscript{3}$ based IR polarizer in reflection geometry.}(a)-(c) show ER as a function of frequency and temperature for thicknesses D1, D2, and D3 respectively. (d)-(f) show DOP as a function of frequency and temperature for thicknesses D1, D2, and D3 respectively.}
\end{figure*}

\begin{figure*}
    \includegraphics[width=\textwidth]{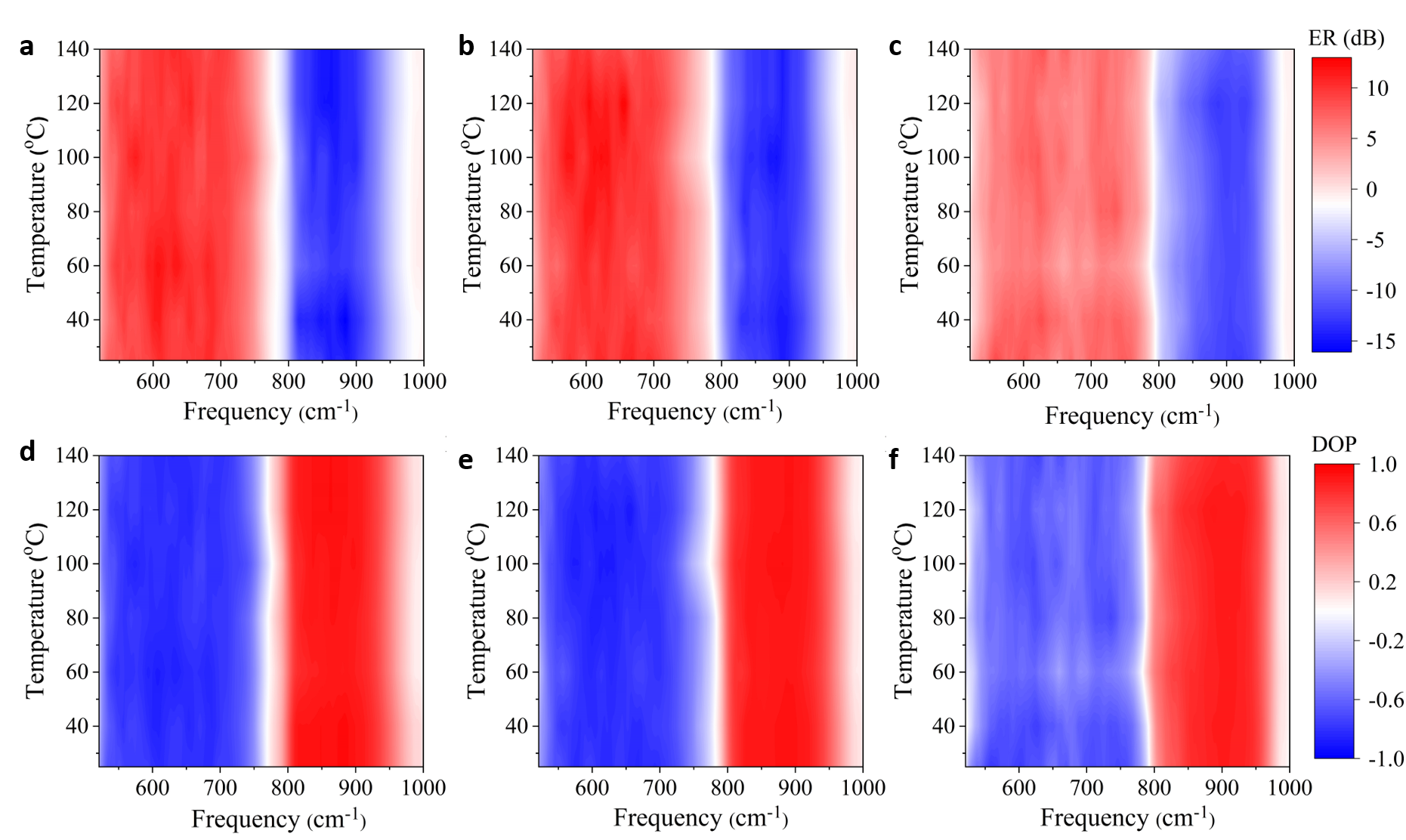}
    \caption{\textbf{Temperature-dependent figures of merit of $\alpha$-MoO$\textsubscript{3}$ based IR polarizer in transmission geometry.}(a)-(c) show ER as a function of frequency and temperature for thicknesses D1, D2 and D3 respectively. (d)-(f) show DOP as a function of frequency and temperature for thicknesses D1, D2 and D3 respectively.}
\end{figure*}

We validate this concept using a finite element method based on full-wave numerical simulation, \textsc{comsol multiphysics}\cite{COMSOL}, for a representative thickness of 2.5$\mu$m. We plot the magnitude of normalized electric field distribution at the frequency 620 cm$\textsuperscript{-1}$ (i.e. RB-1) for incident electric field polarization along [100] and [001] crystal directions of $\alpha$-MoO$_3$ (shown in Fig.~S3 of supplementary information). When the incident electric field is along [001] direction ($s$-polarized), it is majorly reflected by $\alpha$-MoO$_3$, and a small amplitude of the incident light is transmitted. In contrast, when the incident electric field is along [100] direction ($p$-polarized), a small amplitude of the incident electric field is reflected by $\alpha$-MoO$_3$ and majorly transmits the incident electric field. We further develop a transfer matrix method (TMM) based semi-analytical model to estimate the optical responses for $p-$ and $s-$ polarized light incident on $\alpha$-MoO$_3$ as a function of thickness and frequency. Details of the model parameters with reflectance and transmittance color plots are provided in supplementary information Sec.~S1. The findings from TMM corroborate our numerical simulation results. To quantify the performance of $\alpha$-MoO$_3$ based polarizer, we define extinction ratio (ER) as
\begin{align}
    {ER} = {10\log\frac{I_p}{I_s}}
\end{align}
where $I{_p}$ and $I{_s}$ are the reflectances (transmittances) of $p$-polarized and $s$-polarized light respectively. Fig.~3(b)-(c) presents ER as a function of thickness and frequency for reflected and transmitted light, respectively, which are calculated using the TMM approach. Optimum thickness of the $\alpha$-MoO$_3$ based polarizer for large ER and bandwidth is found to be in the range of 2.5 $\mu$m to 3.5 $\mu$m. Corresponding theoretical reflectances and transmittances color plots as a function of frequency and thickness have been shown in Fig. S4 of supplementary information. At higher $\alpha$-MoO$_3$ thicknesses, Fabry-Perot (FP) modes appear for $p$-polarized incident light within RB-1 spectral region\cite{dixit2021mid}. Furthermore, the theoretical transmittance for $s-$ polarized light in RB-1 spectral region is significantly small (~0.001) obtaining which is practically difficult due to noise. Therefore, a reduction of experimental ER is expected in the RB-1 spectral region for transmission as well reflection mode. \par

Fig.~3(d)-(e) represents the ER and DOP, respectively obtained from the experimentally observed polarization-resolved reflectance and transmittance micro-FTIR spectra. These polarization resolved reflectance and transmittance spectra for samples D1 - D3 are shown in Fig.~S4 and Fig.~S5 of supplementary information. It is worth mentioning that experimentally observed reflectance and transmittance spectra show some deviation from their theoretical counterpart (Fig.~S3 of supplementary information). This discrepancy is attributed to multiple factors. Firstly, there are optical losses in the KBr substrate due to contamination from the flake transfer method. Secondly, we obtain a range of thicknesses from the mechanical exfoliation technique. Since FP modes, observed for $p-$ polarized light in RB-1 spectral region, strongly depend on the thickness of $\alpha$-MoO$_3$, range of thickness might lead to modulation of FP modes and consequently the optical responses for $p-$ polarized light in RB-1 spectral region. Hence the theoretical spectrum deviates from the experimental spectrum. We believe that with an improvement in transfer technique, this discrepancy can be resolved and some performance metrics close to the theoretical limit can be obtained. 

To evaluate the operational bandwidth of the proposed IR polarizer, we chose characteristic parameters like ER, so that the experimentally occurred optical losses are compensated in both polarization states of light and give us realistic operational parameters. The ER thresholds for RB-1 and RB-2 spectral regions are considered as 7.5 dB and 10 dB, respectively, to evaluate the operational bandwidth. In the reflection geometry (Fig.~3(d)), a shift of operational spectral bandwidth is observed in the ER spectra towards lower frequency as we go from flake D1 to D3. This is due to the higher thicknesses of $\alpha$-MoO$_3$ from flake D1 to D3, which leads to the change in frequency of FP modes in the RB-1 spectral region resulting in the shift of operational bandwidth in the RB-1 spectral region. Optimum thickness is closest to that of sample D2, which provides a larger operational bandwidth (~ 2.75$\mu$m) compared to  D1 (0.73$\mu$m) and D3 (2.04 $\mu$m) in the RB-1, a trend which is consistent with our theoretical predictions. However, we observe no significant change in the operational bandwidth in the RB-2 spectral region for samples D1 - D3 and operational bandwidth is found around 1.0$\mu$m. Furthermore, the DOP spectrum, given by ${(I_s - I_p)}/{(I_s + I_p)}$, is also shown for D1 - D3 in Fig. 3(d). The DOP informs us about the state of polarization of output light in which $\pm$1 represents completely polarized light, and 0 represents unpolarized light. In the proposed operational bandwidths, DOP is observed to be more than 0.75 and -0.85 in RB-1 and RB-2, respectively, suggesting nearly linear polarized light at the output. In our configuration, $\pm$ signs in DOP correspond to $s-$ and $p-$ polarisation state of output light respectively.

Experimentally measured figures of merit in transmission geometry, displayed in Fig. 3(e), show almost similar bandwidths for samples D1 and D2 for ER values of 7.5~dB and 10~dB in RB-1 and RB-2, respectively. However, ER remains below 7.5~dB in most of RB-1 for D3. This is because sample D3 majorly consists of flakes with thickness around 4.0$\mu$m, which is suboptimal, as discussed earlier. FP modes in the higher thicknesses lead to increased absorption of $p-$ polarized light in RB-1 and result in the reduction of transmittance, which is evident from reflectance and transmittance spectra shown in Fig.~S5 of the supplementary information. Consequently, we observe the reduction of ER for sample D3. Operational bandwidths in transmission geometry are found nearly the same for D1 and D2 in RB-1 (around 4.5 $\mu$m) and RB-2 (around 1.5 $\mu$m) spectral region. These operational bandwidths in RB-1 and RB-2 spectral regions for reflection and transmission geometry have been tabulated in Table-1 for samples D1-D3. Similarly, DOP in the proposed bandwidth for transmission geometry is found more than -0.75 and 0.85 in RB-1 and RB-2, respectively, suggesting nearly linear polarized light at the output. In summary, our FTIR studies confirm that $\alpha$-MoO$_3$ based IR-polarizer, at optimum thickness, exhibits ER of 7.5dB and 10dB in RB-1 and RB-2 spectral region, respectively. The DOP in the proposed operational bandwidth is $\geq$ 0.75 and $\geq$ 0.85 in RB-1 and RB-2, respectively, which suggest nearly linear polarized light at the output. Optimum thickness of $\alpha$-MoO$_3$ for large ER and operational bandwidth is found to be around 2.5 $\mu$m to 3.5 $\mu$m.\par
For the first time to our knowledge, our work explores the above room temperature thermal tolerance of the optical responses of $\alpha$-MoO$\textsubscript{3}$ thin films. Temperature threshold is a critical parameter for optical devices in the mid-IR since the environment and intense IR sources can induce heating and raise the temperature of the polarizer material. In the mid-IR spectral region, optical phonons govern the optical response of $\alpha$-MoO$\textsubscript{3}$ and the lifetime of optical phonons typically decreases with the increase in temperature\cite{silveira2012temperature}. Hence, an enhancement in the optical losses within the material is expected at elevated temperatures, which may lead to deterioration in device performance. Secondly, at high temperatures, the optical phonon frequencies can also shift \cite{Zhu2021}, which may limit the bandwidth of the device. To ensure the temperature stability of $\alpha$-MoO$\textsubscript{3}$ thin film, we first perform temperature-dependent x-ray diffraction up to 160$^{\circ}$C, which is shown in the Fig. S6 of the supplementary information. We found a maximum of around 0.4\% out of plane lattice expansion which is reversible on cooling down to room temperature. Subsequently, we perform temperature-dependent polarization-resolved FTIR reflectance and transmittance investigation of samples D1 - D3 for temperature range from 25$^{\circ}$C to 140$^{\circ}$C. Polarization-resolved reflectance and transmittance of D1-D3 are shown in Figs.~S7 and S8 of supplementary information, respectively. We observe only a slight variation in the optical responses when the temperature is increased from 25$^{\circ}$C to 140$^{\circ}$C, which might be associated with the reduction of phonon lifetimes \cite{silveira2012temperature}. Although the observed intensities are nearly constant within 8\% - 10\% w.r.t. temperature, we have not observed any monotonic trend of intensities. It could be due to the noise resulting from the finite thickness KBr. Furthermore, knife-edge aperture of the FTIR with rough edges might also invoke diffraction-originated noise in the spectrum. These noises mask the expected small variations in the intensities, and hence the presence of any monotonic trend w.r.t. temperature can not be resolved.\par 

Extinction ratio in the reflecting type polarizer as a function of temperature and frequency is shown in Fig.~4(a)-(c) for samples D1 - D3, respectively. We found that ER remains above the threshold values in RB-1 and RB-2 spectral region for the temperature up to 140$^{\circ}$C. Furthermore, DOP for samples D1 - D3 are also found nearly similar to room temperature values for temperature up to 140$^{\circ}$C as shown in Fig.~4(d)-(f). Similarly, ER and DOP for samples D1 - D3 of transmission type polarizer are shown in Fig. 5(a)-(c) and Fig. 5(d)-(f), respectively, where ER for D1 and D2 withhold its threshold up to temperature 140$^{\circ}$C (Fig. 5(a)-(b)) in RB-1 spectral region. Furthermore, the operational bandwidths in both RBs remains almost similar in the entire temperature range. Since sample D3 exhibits ER below threshold even at room temperature due to the higher $\alpha$-MoO$\textsubscript{3}$ thickness, ER of this sample also remains below threshold in most of RB-1 spectral region at higher temperatures as shown in Fig. 5(c). Moreover, DOP in the transmission type polarizers for samples D1 and D2 remains more than -0.75 and 0.85 in RB-1 and RB-2, respectively, up to 140$^{\circ}$C. From these temperature-dependent FTIR investigations, we conclude that $\alpha$-MoO$\textsubscript{3}$ thin film based reflecting and transmission type IR polarizers can retain their figures of merit (ER and DOP) up to 140$^{\circ}$C while retaining similar operational bandwidths in both RBs. Temperature dependent XRD studies confirm that $\alpha$-MoO$\textsubscript{3}$ crystal structure can withstand temperature 140$^{\circ}$C with a reversible 0.4\% of out the plane lattice expansion.\par

\section{Conclusion}
In this work, we investigated the optical properties of large-scale fabricated single crystal $\alpha$-MoO$\textsubscript{3}$ films for application towards mid-IR polarizer in reflection and transmission geometry. Our polarization-resolved Raman studies confirm the inherent in-plane anisotropic nature of the synthesized flakes. Polarization properties of $\alpha$-MoO$\textsubscript{3}$ films in the mid-IR region are attributed to their strong in-plane hyperbolic anisotropy. We found optimum thickness of $\alpha$-MoO$\textsubscript{3}$ films for efficient mid-IR polarizer to be in the range of 2.5$\mu$m to 3.5$\mu$m in which ER is observed to be more than 7.5 dB in RB-1 and more than 10 dB in RB-2 spectral region. Operational bandwidth at optimum thickness is found around 2.75$\mu$m and 1.0$\mu$m in RB-1 and RB-2 spectral region, respectively, for reflection type polarizer and around 4.5$\mu$m and 1.5$\mu$m in RB-1 and RB-2 spectral region, respectively, for transmission type polarizer. To the best of our knowledge, temperature-dependent studies for the optical response of $\alpha$-MoO$\textsubscript{3}$ thin films in the mid-IR have been carried out above room temperature for the first time. This study confirms that our proposed $\alpha$-MoO$\textsubscript{3}$ based polarizer device retains excellent reflectance and transmittance characteristics with a temperature tolerance up to (140$^{\circ}$ C). This work affirms the potential of sub-wavelength vdW thin film of $\alpha$-MoO$\textsubscript{3}$ for miniaturized mid-IR optical devices without using any complicated lithography techniques and can be easily integrated onto the chip-scale platform. This study further opens up the application arena of vdW natural hyperbolic crystals for lithography-free alternatives for other exotic optical functionalities such as quantum interference\cite{nalabothula2020engineering}, sensing\cite{Palermo2020}, planar hyperlensing, and thermal emission control\cite{GomezDiaz2016}.\par

\section{Experimental Section}

\textit{Deposition of $\alpha$-MoO$\textsubscript{3}$ thin films}: Fabrication of $\alpha$-MoO$\textsubscript{3}$ thin films has been carried out here using thermal physical evaporation technique\cite{wang2017growth} in a two-zone split-furnace as shown schematically in the Fig. 1(a). Briefly, 0.1 gm of MoO$\textsubscript{3}$ powder (Sigma-Aldrich) with a purity of 99.99$\%$ has been kept in a cleaned quartz boat at zone-1 (800 \textsuperscript{o}C) and a cleaned silicon substrate is kept around 15 cm away at zone-2 (600 \textsuperscript{o}C) as shown in the Fig.~1(a). Heating rate for both zones is kept around 10 \textsuperscript{o}C per minute. Furnace is kept at target temperature for 120 minutes and then cooled down to room temperature naturally. This process results in the deposition of flakes on a silicon substrate and free-standing thick single-crystal flakes on the wall of the quartz tube. Next, we use the mechanical exfoliation technique of free-standing flakes of $\alpha$-MoO$\textsubscript{3}$ using scotch-tape to obtain the desired thickness of $\alpha$-MoO$\textsubscript{3}$ thin film and transfer it on a KBr window, act like substrate in our case, using thermal release tape.\par

\textit{Structural Characterization of $\alpha$-MoO$\textsubscript{3}$ thin films}: HR-TEM and SAED pattern was obtained by FEI Tecnai G2, F30 with an acceleration voltage of 300 kV. Temperature-dependent structural properties of $\alpha$-MoO$\textsubscript{3}$ thin films were studied from Rigaku X-ray diffractometer using Cu- K-$\alpha$ radiation (1.541 $\text{\AA}$), operated at 40 kV of potential and 40 mA of current, in the $2\theta-\omega$ mode with a scan rate of 6 degrees per minute.The samples were heated at a rate of 10 $^{\circ}$ C/min with 10 min of stay time at target temperature. Morphology of $\alpha$-MoO$\textsubscript{3}$ thin-films on silicon substrate was observed using JEOL-JSM scanning electron microscope operated at an accelerating voltage of 20kV in the secondary electron mode. X-ray photoelectron spectroscopy (XPS) was carried out using Kratos Analytical, Axis Supra for elemental composition analysis. The spectrometer was calibrated with respect to the binding energy of 1s of carbon at 285.0 eV. Thickness profiles of various deposited and transferred flakes were investigated usingn Bruker's DekTak XT stylus profilometer. Polarization-resolved Raman spectroscopy was carried out in back-scattering geometry using Horiba’s Jobin Yvon Lab-Ram 800 at room temperature. A 532 nm solid-state laser line was used as an excitation source and a 100x long working distance objective with a numerical aperture (NA) of 1.25 was used for focusing the laser beam to a spot size of around 0.5 $\mu$m. Raman spectrometer was initially calibrated using Raman active optical mode of crystalline silicon at 520.70 cm$\textsuperscript{-1}$. An analyzer is placed in the path of back-scattered light before the spectrometer’s entrance, which allows us to analyze the Raman shift parallel or perpendicular to incident laser light. Rotation of flake method is adopted using rotational stage keeping the polarization state of incident laser light fixed.\par

\textit{Fourier-Transform Infrared Spectroscopy}: Mid-infrared optical response of the $\alpha$-MoO$\textsubscript{3}$ thin films were studied in the 520 cm$\textsuperscript{-1}$ -- 1000 cm$\textsuperscript{-1}$ spectral region using Bruker Hyperion-3000 FTIR microscope equipped with Bruker Vertex 80 spectrometer. Temperature-dependent studies were carried out upto 140$^{\circ}$C with an interval of 20$^{\circ}$C using a Hyperion heatable sample holder, and a Pt100 resistor was connected for sensing the current temperature. Temperature control unit uses a digital PID algorithm for controlling the temperature. Stay time at the target temperature was kept at 10 min. The IR light is illuminated using a 15X Cassegrain objective with a NA of 0.40 and an average off-normal incident angle of about 17$^{\circ}$. Spectrum is recorded with the help of a wideband MCT (mercury cadmium telluride) detector at the spectral resolution of 2 cm\textsuperscript{-1} and 256 number of scan. Flakes are kept in such a manner that the x-axis of the FTIR instrument is parallel to the [100] direction of $\alpha$-MoO$\textsubscript{3}$. The polarization-dependent studies were carried out using a KRS-5 based IR polarizer. Here, reflectance spectra are collected with respect to the gold background in the back-scattering geometry, and transmittance spectra are collected with respect to the air background. A knife-edge aperture is set with an area 90$\mu$m X 130$\mu$m for the selection of area on $\alpha$-MoO$\textsubscript{3}$ flake. All FTIR spectra are smoothened by weighted adjacent averaging from 9 points to remove noises.

\emph{Acknowledgement.}  N.R.S. acknowledges the Council of Scientific $\&$ Industrial Research fellowship No: 09/087(0997)/2019-EMR-I. S.D. acknowledges financial support from Institute Postdoctoral Fellowship IIT Bombay. A.K.S. acknowledges Industrial Research and Consultancy Center for the financial support. A.K. acknowledges funding from the Department of Science and Technology via the grants: SB/S2/RJN-110/2017, ECR/2018/001485 and DST/NM/NS-2018/49. We acknowledge Industrial Research and Consultancy Center and the Center for Research in Nanotechnology and Science, IIT Bombay for providing access to characterization facilities.

\addcontentsline{toc}{section}{Bibliography}
\bibliographystyle{unsrt}
\bibliography{ref.bib}

\end{document}